\documentclass[twocolumn,tighten]{aastex6}

\newcommand\chandra{{\it Chandra}}
\newcommand\xmm{{\it XMM-Newton}}
\newcommand\swift{{\it Swift}}
\newcommand\hst{{\it HST}}
\newcommand\rosat{{\it ROSAT}}
\newcommand\einstein{{\it Einstein}}

\newcommand\ctssec{{\rm~count\ s}^{-1}}
\newcommand\ergsec{{\rm~erg\ s}^{-1}}
\newcommand\cmsqre{{\rm~cm}^{-2}}
\newcommand\Msun{\rm M_{\odot}}

\begin{document}

\title{Transient X-ray Sources in the Magellanic-type Galaxy NGC 4449}
\shorttitle{Transients in NGC 4449}

\author{V. Jithesh and Zhongxiang Wang}
\affil{Shanghai Astronomical Observatory, Chinese Academy of Sciences, 80 Nandan Road, Shanghai 200030, China; jithesh@shao.ac.cn}
\shortauthors{Jithesh \& Wang}

\begin{abstract}
We report the identification of seven transient X-ray sources in the 
nearby Magellanic-type galaxy NGC 4449 using the archival multi-epoch 
X-ray observations conducted with {\it Chandra}, {\it XMM-Newton} and 
{\it Swift} telescopes over year 2001--2013. Among them, two sources 
are classified as supersoft X-ray sources (SSSs) because of their soft X-ray 
color and rest of the sources are X-ray binaries (XRBs). 
Transient SSSs spectra can be fitted with a blackbody 
of effective temperature $\sim 80-105$ eV and luminosities were
$\simeq 10^{37} - 10^{38} {\rm~erg\ s}^{-1}$ in 0.3--8 keV. These properties
are consistent with the widely accepted model for SSSs,
an accreting white dwarf with the steady nuclear burning on 
its surface, while the SSS emission has also been observed in many post-nova systems. 
Detailed analysis of one sufficiently bright SSS 
revealed the strong short-term variability, possibly showing a 2.3 hour 
periodic modulation, and long-term variability, detectable over 23 years with 
different X-ray telescopes before year 2003. The X-ray properties of four other 
transients are consistent with neutron star or black hole binaries in their hard state, 
while the remaining source is most likely an XRB with a quasi-soft X-ray spectrum. 
Analysis of archival {\it Hubble Space Telescope} image
data was also conducted, and multiple massive stars were found as
possible counterparts. We conclude that the X-ray transient properties in
NGC 4449 are similar to those in other Magellanic-type galaxies.
\end{abstract}


\section{Introduction} 
\label{sec:intro}

NGC 4449 is a Magellanic-type irregular star-forming galaxy, first studied at
X-ray wavelengths with the \einstein{} satellite \citep{Fab92}. 
The \einstein{} IPC and HRI observations detected three point sources in 
the galaxy and one of them coincides with an extremely luminous 
supernova remnant (SNR), previously identified with radio observations \citep{Sea78}. 
The \rosat{} HRI observations resolved seven luminous X-ray sources in 
the $\rm D_{25}$ region of NGC 4449 and identified an extended emission feature, which 
is mainly from the unresolved point sources and hot interstellar medium \citep{Vog97}. 
Later, \chandra{} observation (conducted in 2001) has detected 24 X-ray point 
sources in the optical extend of NGC 4449 and the X-ray point source population 
has been determined to be dominated by X-ray binaries (XRBs) along with 
supersoft X-ray sources (SSSs) and SNRs \citep{Sum03}. It has been found 
that the X-ray emission from the Magellanic-type 
galaxies dominantly comes from young population objects and the X-ray sources 
are spatially associated with the star forming 
regions \citep[e.g,][and references 
therein]{Sto06}. The case of NGC 4449 appears to be consistent with this 
feature, as the bright XRBs are associated with young star 
clusters \citep{Ran11}. 

The transient X-ray source population has been identified in Magellanic-type 
galaxies such as the Small and Large Magellanic Clouds (SMC and LMC) and NGC 55 
\citep[][and references therein]{Kah96, Coe01, Jit16}. The majority of 
the transient sources are XRBs, including low-mass and high-mass XRBs. 
In addition, another important class are SSSs, which contribute approximately 
$25 -35 \%$ of the total transient 
population and are the inevitable class of objects in these galaxies. 

SSSs are soft X-ray emitting objects, with
spectra resembling a blackbody in a temperature range of 20 -- 100 eV and
X-ray luminosities being in a range of $10^{35} - 10^{38} {\rm~erg\ s}^{-1}$ 
\citep[see reviews in] {Kah97, Dis10}. SSSs were first discovered with 
\einstein{} observatory \citep{Lon81}, and \rosat{} later discovered more of
them in Magellanic Clouds (MCs) and in our Galaxy \citep{Tru91, Gre91, Beu95}, 
establishing them as an important new class of X-ray objects. Supported by many 
observational studies, the steady thermonuclear burning of accreted matter on 
the surface of a white dwarf \citep[WD;][]{Van92} is the most accepted explanation for SSSs. 
In addition, SSS emission has also been observed in many nova systems 
\citep[][and references therein]{Oeg84, Hen10, Hen11, Hen14}. For these cases, 
a fraction of the hot envelope, most of which has been ejected during a nova 
explosion, still remains on the surface of a WD \citep{Sta74} and powers SSS emission. 
Typical SSSs have little emission above 1 keV, while 
a companion class of SSSs has been identified in the external galaxies 
that have X-ray emission above 1 keV. The latter systems are referred as 
quasi-soft sources (QSSs) with temperature in a range of $100-350$ eV 
\citep{Dis04}. Several SSSs and QSSs identified in the external galaxies 
\citep{Gre00, Dis03, Dis04} are associated with diverse types of objects, 
which include XRBs with a WD, a neutron star (NS), or a stellar-mass 
black hole (BH) as the compact object, ultra-luminous SSSs \citep{Kon04, Liu08} 
possibly consisting of an intermediate-mass BH, and SNRs \citep{Ori06}. The 
identification studies have further expanded the SSS class.

In this paper, we re-examined the X-ray observations of the Magellanic-type 
galaxy NGC 4449 to search for the transient X-ray sources. In the following, 
\S 2 describes the observations and data reduction processes used. 
We explain the analysis and results in \S 3. In \S 4, 
we discuss and conclude the properties of the transient X-ray sources
found in NGC 4449.

\section{Observations and Data Reduction}
\subsection{\chandra{} Observations}
\label{sec:chandra}

We found four sets of archival \chandra{} observations of NGC 4449. The details 
of the observations are given in Table \ref{obslog}. The \chandra{} 
observation conducted in 2000 October (ObsID: 938) 
pointed at a target which is $\sim 17$ arcmin away from NGC 4449 and 
the sources considered in this 
analysis were not covered by this observation. Hence we did not include this 
data in the analysis. The data from the other three observations were taken 
with the Advanced CCD Imaging Spectrometer Spectroscopy Array (ACIS-S), and
NGC 4449 was positioned on the back-illuminated S3 chip. We processed the data 
using the \chandra{} Interactive Analysis of Observation (CIAO) software 
(version 4.6) and \chandra{} Calibration Data Base (CALDB) 
version 4.6.1.1. The source detection routine was carried out in the ACIS data 
over the 0.3 -- 8 keV energy band. We used CIAO's Mexican-hat wavelet source 
detection routine, {\sc wavdetect} \citep{Fre02}, to detect the X-ray sources 
in NGC 4449. After the production of an exposure map, we ran 
the {\sc wavdetect} tool with wavelet scales of 1.4, 2, 4, 8 and 16 pixels 
and a detection threshold of $5\times 10^{-7}$. 

\begin{table}
\tabletypesize{\small}
\tablecolumns{5}
\setlength{\tabcolsep}{4.0pt}
\tablewidth{320pt}
	\caption{Observation log}
 	\begin{tabular}{@{}lcccr@{}}
	\hline
	\hline
\colhead{Observatory} & \colhead{Data} & \colhead{ObsID} & \colhead{Date} & \colhead{Exposure$^a$} \\
\hline


		\chandra{} & C0 & 938 & 2000 Oct 04 & 5.7 \\
			   & C1 & 2031 & 2001 Feb 04 & 26.9 \\
			   & C2 & 10125 & 2009 Mar 04 & 15.1 \\
			   & C3 & 10875 & 2009 Mar 07 & 60.2 \\
		{\it XMM-} & XMM1 & K601 & 2002 May 25 & 23.9 \\
		{\it Newton} & XMM2 & K701 & 2002 Jun 02 & 15.9 \\
		{\it Swift} & S1 & L3001 & 2007 Mar 27 & 4.7 \\ 
			    & S2 & M1007 & 2013 Dec 06 & 4.9 \\			    
		\hst{}     & POS-A  & 10585 & 2005 Nov 10 & 3.7 (F435W) \\
			   & 	    &       &             & 2.5 (F555W) \\
		   	   & 	    &       &             & 2.1 (F814W) \\
		   	   & POS-B  & 10585 & 2005 Nov 11 & 3.5 (F435W) \\
			   & 	    &       &             & 2.5 (F555W) \\
		   	   & 	    &       &             & 2.1 (F814W) \\
			   & POS-C  & 10585 & 2005 Nov 17 & 0.5 (F814W) \\ 
\hline
\end{tabular} 
\tablecomments {$^{a}$Exposure time is in unit of kilo seconds. The prefix K, L and M denote 0112521, 0003587 and 0008226 respectively.}
\label{obslog}
\end{table}

\begin{table*}
\tabletypesize{\small}
\tablecolumns{8}
\setlength{\tabcolsep}{5.0pt}
\tablewidth{320pt}
	\caption{Transient X-ray sources in NGC 4449}
 	\begin{tabular}{@{}lccccccr@{}}
	\hline
	\hline
\colhead{Src} & \colhead{ObsID} & \colhead{Catalog} & \colhead{R.A.} & \colhead{Decl.} & \colhead{SC} & \colhead{HC} & \colhead{Class} \\
\colhead{No.}    &                 &  \colhead{No. \& Ref.} & \colhead{(h:m:s)} & \colhead{(${^\circ}:':''$)} &  & & \\
\hline


		TSS1 & C1   & X3(AV97), 14(LKS03) & 12:28:10.94 & +44:03:38.18 & $-0.99\pm0.01$ & -- & SSS \\
		     & XMM1-PN &   		  & 	     	& 	       & $-0.44\pm0.09$ & -- & SSS \\
		     & XMM1-MOS &   		  & 	     	& 	       & $-1.18\pm0.34$ & -- & SSS \\		     
		     & XMM2-PN &   		  & 	     	& 	       & $-0.99\pm0.02$ & $0.33\pm0.53$ & SSS \\
		     & XMM2-MOS &   		  & 	     	& 	       & $-0.96\pm0.06$ & $-0.99\pm1.99$ & SSS \\		     
		TSS2 & C1 & 28(LKS03), P18(JO05), X19(BR11) & 12:28:19.03 & +44:05:44.59 & $-0.96\pm0.04$ & -- & SSS \\
		T3 & C1 & 6(LKS03), X13(BR11) & 12:28:06.88 & +44:05:28.27 & $0.25\pm0.08$ & $-0.67\pm0.20$ & XRB \\
		T4 & C1 & 2(LKS03), X8(BR11) & 12:28:00.73 & +44:04:32.24 & $0.30\pm0.12$ & $0.03\pm0.01$ & XRB \\
		
		T5 & C3 & X43(BR11) & 12:28:20.03 & +44:06:21.36 & $1.00\pm0.44$ & $0.14\pm0.04$ & ABS \\
		T6 & C3 & X44(BR11) & 12:28:13.25 & +44:06:46.05 & $0.09\pm0.06$ & $0.29\pm0.10$ & XRB \\
		T7 & C3 & X46(BR11) & 12:28:09.71 & +44:05:19.35 & $-0.44\pm0.08$ & $-0.67\pm0.24$ & SNR \\

\hline
\end{tabular} 
\tablecomments {(1) Source number used in this paper; (2) observation ID in which the source was detected; (3) source identification number from previous studies, AV97=\citet{Vog97}, LKS03=\citet{Sum03}, JO05=\citet{Ott05}, BR11=\citet{Ran11}; (4)-(5) right ascension (R.A.) and declination (Decl.) of each source (J2000.0); (6)-(7) X-ray colors derived from the count rate (See \S \ref{sec:trans}); (8) source class according to the classification scheme of \citet{Kil05} and \citet{Jen05}.}
\label{color}
\end{table*}

\subsection{\xmm{} and \swift{} Observations}
\label{sec:xmm}

NGC 4449 was observed twice with the \xmm{} telescope \citep{Jan01}. 
The European Photon Imaging Camera (EPIC) PN \citep{Str01} and metal oxide 
semiconductor \citep[MOS;][]{Tur01} camera were operated in the full frame 
mode and the thin filter was used in both observations. 
The details of the \xmm{} observations of NGC 4449 are given in 
Table \ref{obslog}. The data sets were processed with the Science Analysis 
Software (SAS version 14.0). We processed PN and MOS data using {\sc epchain} 
and {\sc emchain} to produce calibrated photon event files, and used 
unflagged single and double pixel events with 
{\sc pattern} 0--4 and 0--12 to filter the processed PN and MOS events 
respectively. To exclude particle flaring background, we created a 
Good Time Interval (GTI) file above 10 keV for the full field 
using the task {\sc tabgtigen}. The general cut-off value was 
RATE $< 0.4\ctssec$. However for the XMM1 observation
(Table~\ref{obslog}), which has a high particle background, we 
used much higher cut-off value (RATE $< 4\ctssec$) for the filtering. 
As a result, we obtained $\sim 1$ ks usable data from this observation. 
To detect the X-ray sources from \xmm{} observations, 
we ran the source detection routine available in SAS, {\sc edetect\_chain}, 
using the standard parameters for 
EPIC-PN data over the entire energy band as well as the $0.3-1$ keV, 
$1-2$ keV, and $2-8$ keV bands. 

In addition, we checked the long-exposure ($>$ 4 ks; see Table \ref{obslog}) 
observations of the target field conducted with the \swift{} X-ray 
Telescope \citep[XRT;][]{Bur05} for possible 
detection of transient sources found in the \chandra{} and \xmm{} observations.
However, none of the transient sources were detected 
in the \swift{} XRT observations. 

\subsection{{\it Hubble Space Telescope} Observations}

We searched for possible optical counterparts of transient X-ray sources in 
the {\it Hubble Space Telescope} (\hst{}) observations. We used image data
taken with the Wide Field Channel (WFC) camera of the Advanced Camera for 
Surveys (ACS), which were downloaded from the Hubble Legacy Archive\footnote{http://hla.stsci.edu/} (HLA). 
The images were generated from the individual flat-fielded exposures using 
the IRAF task {\sc multidrizzle\footnote{http://www.stsci.edu/hst/HST\_overview/documents/multidrizzle}}. 
Three sets of the images covering NGC 4449 (POS-A, POS-B and POS-C) 
were taken with the F435W ($B$), F555W ($V$), 
and F814W ($I$) filters (see Table \ref{obslog} for details).

\section{Analysis and Results}

\subsection{Transients and their X-ray Colors}
\label{sec:trans}

To search for transient sources, we mainly used the long-separated 
and long-exposure \chandra{} observations (C1 and C3), while the other 
\chandra{} and \xmm{} observations (C2, XMM1, and XMM2) were used for 
checking the luminous fast transients and follow-up studies. We detected 24 and 23 X-ray 
sources in the $\rm D_{25}$ region of the galaxy in the 2001 (C1) and 
2009 (C3) \chandra{} observations respectively. Four sources detected 
in C1 did not appear in C3, and on the other hand three sources in C1 
were brighter in C3. However, no other transient events were 
detected in the short-exposure observations. Thus, we identified seven 
transient X-ray sources in NGC 4449 (See Table \ref{color}). We ran the source 
detection algorithm again on the soft ($S$: $0.3-1$ keV), 
medium ($M$: $1-2$ keV), and hard ($H$: $2-8$ keV) images 
and the transient sources were detected in at least one of the three energy 
bands.

The count rates of the transient sources in the three energy bands 
were obtained, and we calculated two 
X-ray colors, defined as, SC=$(M-S)/(M+S)$, HC=$(H-M)/(H+M)$. 
To understand the source class of the transient sources, we compared 
the X-ray colors of the sources 
with the color classification scheme provided in \cite{Kil05} 
and \cite{Jen05} for \chandra{} and \xmm{} sources respectively. 
For the \xmm{} observations, we defined the $0.3-1$ keV, $1-2$ keV, 
and $2-6$ keV energy band as $S$, $M$, and $H$, respectively. 
The transient source positions, X-ray colors, and their respective source 
classes are given in Table \ref{color}. Out of seven sources, 
three of them (T3, T4 and T6) are in the XRB class, one 
source each in the absorbed (ABS) and SNR category (T5 and T7 respectively), 
and the rest of the two sources, named as TSS1 and TSS2, belong to 
the SSS class. Earlier studies \citep{Vog97, Sum03, Ott05, Ran11} also identified 
these sources in their respective classes. Here we further 
classified them as the transient X-ray sources in NGC 4449. 
Among them, T7 has a color consistent with SNR, but the transient behavior 
rules out the SNR nature of this source. In the following sections, we describe 
their detailed properties we obtained.

\subsection{TSS1} 

TSS1 is one of the brightest sources in NGC 4449 and was also detected by 
previous X-ray telescopes \citep[\einstein{} and \rosat{};][]{Fab92, Vog97}. 
The source was detected in the C1 observation \citep{Sum03} but was not in 
the C2 and C3 observations. For the detection, we extracted the source 
and background events within a circular region with a radius of 5 arcsec. 
Using the CIAO {\sc specextract} tool, we obtained the source and background 
spectra along with the ancillary response file (ARF) and redistribution matrix 
file (RMF). The source spectrum was grouped with a minimum of 20 counts per bin and 
the spectral analysis was performed in the $0.3-8$ keV energy range with {\tt XSPEC} 
version 12.8.1g \citep{Arn96} available in the HEASOFT (version 6.15.1). 
TSS1 has a few tens of spectral counts below 0.3 keV energy. Since the calibration of 
the spectral response of the instruments is less certain at the low energies, we did not 
include these photons in the spectral analysis. However, they were included in the light curve 
analysis (see \S \ref{long_lc}). In the undetected observations, we 
computed the upper limits on the count rate ($90\%$ confidence level) 
using the {\sc aprates} task in CIAO. 

In both \xmm{} observations, the source was detected and we used a 
circular region with a radius of 18 arcsec for both source and background 
event extraction. For the EPIC-PN and MOS data, 
the source and background spectra along with associated files (ARF and RMF) 
were extracted using the SAS tool {\sc especget}, and fitted simultaneously. 

We used a blackbody ({\tt BBODY}) model to fit the spectra, which was modified 
by the Galactic absorption, $N_{\rm H, Gal}= 1.61\times10^{20}\cmsqre$ \citep{Kal05}. 
For all the spectra, the absorption-corrected fluxes were derived in the $0.3-8$ keV 
energy band using the convolution model {\tt CFLUX} available in {\sc xspec}, 
and the luminosities were calculated by assuming a distance of 2.93\,Mpc \citep{Kar98}.
All errors quoted were computed at a 90\% confidence level. The best-fit 
spectral parameters are given in Table \ref{sss}. 
We tested to add another absorption component ({\tt tbabs}), which was considered 
to account for the local absorption within the galaxy and/or material around the source. 
The additional absorption component did not improve the spectral fit compared to 
the absorbed blackbody model and the absorption values were not well constrained, 
$N_{\rm H} < 10.4\times10^{20}\cmsqre$.

\begin{table}
\tabletypesize{\small}
\tablecolumns{5}
\setlength{\tabcolsep}{5.0pt}
\tablewidth{320pt}
	\caption{Spectral parameters for transient X-ray sources in NGC 4449}
 	\begin{tabular}{@{}lccccr@{}}
	\hline
	\hline
\colhead{Src} & \colhead{ObsID} & \colhead{kT/$\Gamma$} & \colhead{log $L_{\rm X}$} & \colhead{$\chi^2/\rm d.o.f$} & \colhead{$G$}\\
\colhead{No.}    &                 &                    & \colhead{$\ergsec$} & & \\
\hline


		TSS1 & C1   & $104.9^{+10.3}_{-8.9}$ & $37.87^{+0.04}_{-0.04}$ & 9.3/12 & 0.68 \\
		     & XMM1 & $89.3^{+22.0}_{-20.8}$ & $38.11^{+0.14}_{-0.19}$ & 26.3/19 & 0.12 \\
		     & XMM2 & $86.0^{+6.5}_{-6.1}$ & $37.85^{+0.04}_{-0.05}$ & 26.6/21 & 0.19 \\
		     & C2   &       ...              & $<36.51$                &  ...  &  ... \\
		     & C3   &       ...              & $<36.25$        	       &  ...  & ... \\
		     
		TSS2 & C1   & $76.1^{+16.0}_{-13.7}$ & $36.92^{+0.13}_{-0.15}$ & 9.1/7(C) & 46\% \\
		     & XMM1 & 		...	     & $<37.24$		       & ...	& ...\\
		     & XMM2 & 		...	     & $<36.70$		       & ...	& ...\\
		     & C2   &       ...              & $<36.37$                & ...    & ...\\
		     & C3   &       ...              & $<36.05$                & ...    & ...\\
		     
		T3 & C1   & $1.36^{+0.95}_{-1.35}$ & $36.71^{+0.29}_{-0.32}$ & 3.9/3(C) & 26\%\\
		   & C2   & 		...	   & $<36.95$		       & ...	& ...\\
		   & C3   & 		...	   & $<36.59$		       & ...	& ...\\
		   
		T4 & C1   & $<2.35$                & $37.09^{+1.29}_{-0.80}$ & 0.2/2(C) & 1\% \\
		   & C2   & 		...	   & $<36.21$		       & ...	& ...\\
		   & C3   & 		...	   & $<36.28$		       & ...	& ...\\
	
		T5 & C1   & 		...	   & $<36.74$		       & ...	& ...\\
		   & C2   & 		...	   & $<36.63$		       & ...	& ...\\
		   & C3   & $0.38^{+0.73}_{-0.80}$ & $36.89^{+0.27}_{-0.27}$ & 2.3/4(C) & 10\% \\

		T6 & C1   & 		...	   & $<36.85$		       & ...	& ...\\
		   & C2   & 		...	   & $<36.97$		       & ...	& ...\\
		   & C3   & $0.88^{+0.93}_{-0.80}$ & $36.86^{+0.21}_{-0.21}$ & 8.2/4(C) & 56\% \\

		T7 & C1   & 		...	   & $<36.59$		       & ...	& ...\\
		   & C2   & 		...	   & $<36.82$		       & ...	& ...\\
		   & C3   & $3.65^{+1.03}_{-0.96}$ & $36.94^{+0.28}_{-0.27}$ & 12.5/5(C) & 70\% \\
   
\hline
\end{tabular} 
\tablecomments {The absorption was fixed at Galactic value in all cases. (1) Source number; (2) observation ID used in each fit; (3) blackbody temperature in eV for TSS1 and TSS2 and photon index for rest of the sources; (4) absorption-corrected X-ray luminosity in the $0.3-8$ keV energy band; (5) the $\chi^2/\rm d.o.f$ value for the model and `C' in the bracket indicates that C-statistics is used for spectral modeling; (6) goodness-of-fit (Null hypothesis probability for $\chi^2$ statistics and ``Goodness'' value when using C-Statistics).}
\label{sss}
\end{table}

\subsubsection{Light Curve and Long-term Variability}
\label{long_lc}

\begin{figure}
\includegraphics[width=\columnwidth]{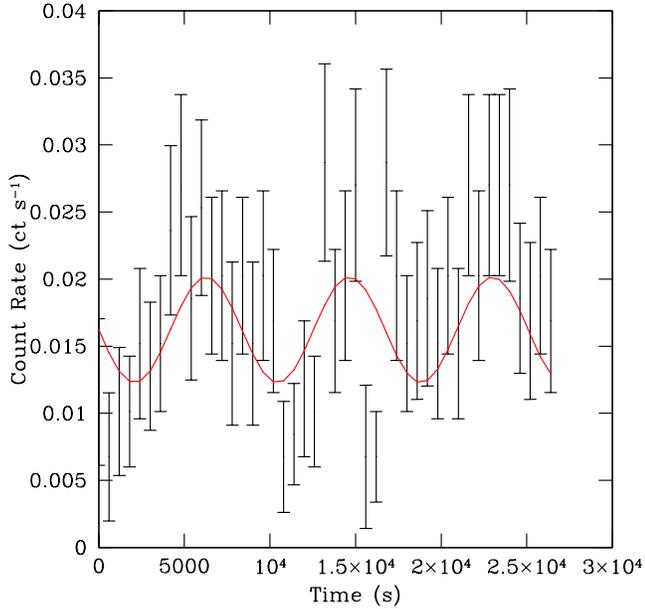}
\caption{Background subtracted ACIS-S (0.2 -- 2 keV) light curve of TSS1 
with a binsize of 600 s. A sinusoidal modulation with period of 139 min 
fitted to the data is also shown.}
\label{lctss1}
\end{figure}

We investigated the source variability using the results from the
\chandra{} and \xmm{} observations. We derived the binned and background 
subtracted light curve using CIAO task {\sc dmextract}. 
For the light curve, the extraction regions were the same as those for obtaining
the spectra. Since low-energy emission from the source is dominant, we 
only included the events in the 0.2 -- 2 keV 
energy band in the light curve. The light curve of TSS1 obtained from the 
C1 observation is shown in Figure \ref{lctss1}. There is clear variability 
seen in the light curve. We tested to fit a sinusoid to the light curve, 
which gave a reduced $\chi^2$ of 1.2 for 42 
degrees of freedom (d.o.f). While the low $\chi^2$ value is due to the large
uncertainties of the data points, a period of $139\pm4$ min was found. 
This period is also marginally consistent with a peak ($\sim 160$ min) obtained 
from the power spectral analysis (using {\tt powspec} task). 

In the \xmm{} observations, the source and background light curves were 
extracted from the cleaned event files and corrected using the SAS 
task {\sc epiclccorr}. The quality and length of the light curve 
do not allow us to confirm the periodic variability seen in the \chandra{} 
observation. However, we investigated the short-term X-ray variability 
by performing a Kolmogorov-Smirnov (K-S) test on the light curve. From 
the K-S test, we found that the source showed a strong X-ray variability 
in short time scale (600 s) at a confidence level of $> 99\%$.

\begin{figure}
\includegraphics[width=\columnwidth]{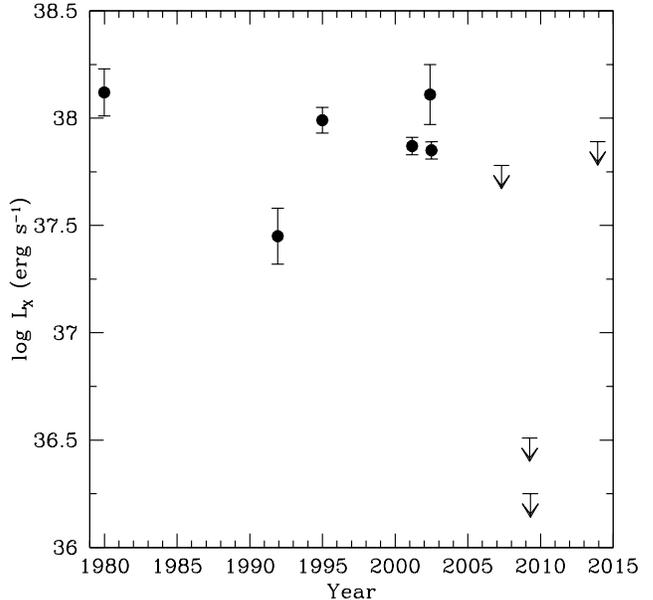}
\caption{Long-term $0.3-8$ keV light curve of TSS1 based on \einstein{}, \rosat{}, \chandra{}, \xmm{} and \swift{} 
observations. See the text for details.}
\label{longlc}
\end{figure}

We also studied the long-term variability of TSS1 using \chandra{}, \xmm{} and 
\swift{} observations. The baseline of the long-term 
light curve was extended by adding the data points from \einstein{} HRI, 
\rosat{} PSPC, and 
\rosat{} HRI observations. NGC 4449 was observed with HRI 
onboard \einstein{} observatory in December 1979 \citep{Fab92} and 
TSS1 was identified as one of the three X-ray sources in the galaxy
(HEASARC HRIEXO catalog\footnote{http://heasarc.gsfc.nasa.gov/W3Browse/einstein/hriexo.html}). 
The source was bright during the \einstein{} observation with a count rate 
of $(1.1\pm0.3)\times 10^{-3}\ctssec$. The \rosat{} PSPC and HRI observations
were conducted in November 1991 and December 1994 respectively \citep{Vog97}, 
and TSS1 (X3 in \rosat{}) was detected in both observations with count rates of 
$(3.3\pm1.1)\times 10^{-3}$ and $(27.3\pm4.4)\times 10^{-4}\ctssec$ 
respectively. In the available long-exposure \swift{} XRT observations, 
TSS1 was not detected and we derived the upper limits of the count rates 
using the {\tt uplimit} command in XIMAGE. 
We converted the \einstein{}, \rosat{} and \swift{} XRT count rates into 
fluxes in the $0.3-8$ keV energy band using the webPIMMS\footnote{http://heasarc.gsfc.nasa.gov/cgi-bin/Tools/w3pimms/w3pimms.pl} 
by assuming an absorbed blackbody model. Figure \ref{longlc} shows 
the long-term light curve of TSS1 over $\sim 34$ years.
The source appeared to have a constant X-ray luminosity 
($\sim 10^{38}\ergsec$) over a period of $\sim 23$ years, 
although the ill-sampled observations could have missed the intervals of 
stronger or weaker states of the source. The source then
became undetectable in the \swift{} (conducted in 2007) and two \chandra{} 
observations conducted in 2009, which indicate a flux change of at least 
two orders of magnitude. We estimated the duty cycle of TSS1 as the ratio of the time spent 
in ``on-state'' (observations where the source was detected) to 
the total time in both ``on-state'' and ``off-state'' (the non-detection 
observations). Assuming that TSS1 was in the on-state during $1979-2002$ 
period, the duty cycle is $\sim$0.6.

\subsection{TSS2}     

TSS2 (CXOJ122819.03+440544.59) was detected only in the C1 observation. 
We extracted the source and background events from a circular region with
a radius of 3 arcsec. The extracted spectrum was fitted with an absorbed 
blackbody model, and the best-fit spectral parameters 
are $kT=76.1^{+16.0}_{-13.7}$ eV 
with $L_{X}=(8.3^{+2.9}_{-2.4})\times10^{36}\ergsec$ (see also Table \ref{sss}).
Since TSS2 has limited net spectral counts, we adopted 
Cash-Statistics \citep{Cas79} for the spectral modeling and the quality 
of the fit was obtained by performing 5000 Monte Carlo simulations 
using {\tt goodness} task. The source was then not detected in the XMM1, 
XMM2, C2 and C3 observations. We derived 
the upper limits on the count rate (90\% confidence level) 
from a circular region of 10 arcsec radius using the SAS task 
{\sc eregionanalyse} for the \xmm{} observations. The upper limits were 
converted into fluxes by assuming the best-fit 
model parameters obtained from the C1 observation. The upper limits are 
provided in Table \ref{sss}. We extracted the background-subtracted 
light curve of TSS2, binned with 100 s and 600 s intervals, and 
performed the K-S test to investigate the variability. However, the limited 
net counts do not allow us to obtain any meaningful results from 
the light curves. 

\subsection{Other Transient Sources} 

The five transient sources (T3, T4, T5, T6 and T7) were detected either 
in C1 or C3 observations. We extracted the spectra and upper limits of 
these sources using CIAO tools. Since these sources have limited net 
counts, a detailed analysis is not possible. 
We tested to fit their spectra with absorbed power law (PL) model 
and the results are presented in Table \ref{sss}. This model gives a 
reasonable fit to all sources, except T7. T7 has a steep PL index 
($\sim 3.7$) from the absorbed PL fit. Thus, we tested {\tt BBODY} model 
for T7 and this model marginally improved the spectral fit 
(in terms of goodness-of-fit) over the PL model. The spectral parameters 
for {\tt BBODY} model are $kT= 177.4^{+45.6}_{-33.2}$ eV, 
$L_{X}=(4.2^{+1.8}_{-1.4})\times10^{36}\ergsec$ with $\chi^2/\rm d.o.f = 9.5/5$. 
In the \xmm{} observations, these sources were not well resolved and 
undetected. In some cases, they were located in the edges of the CCD 
and contaminated by other nearby sources. 
Hence, the \xmm{} upper limits of these sources were not obtained. 
The deep \chandra{} observations provided the upper limits in all cases.

\subsection{Search for Possible Optical Counterparts} 
\label{counterpart}

\begin{figure}
\centering
\begin{tabular}{cc}
 TSS1 & TSS2  \\
\includegraphics[width=3.5cm,height=3.5cm,angle=0]{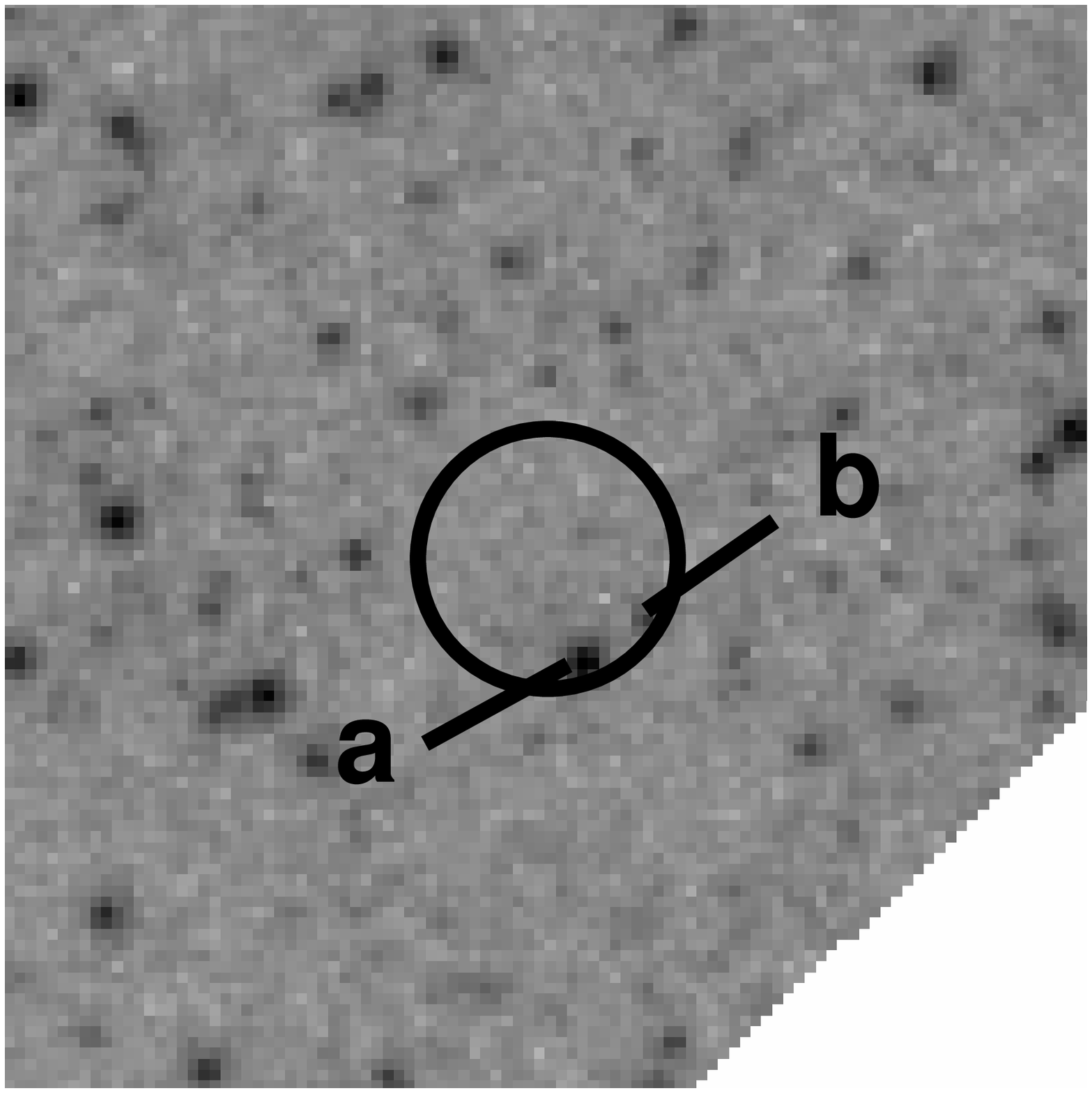} & 
\includegraphics[width=3.5cm,height=3.5cm,angle=0]{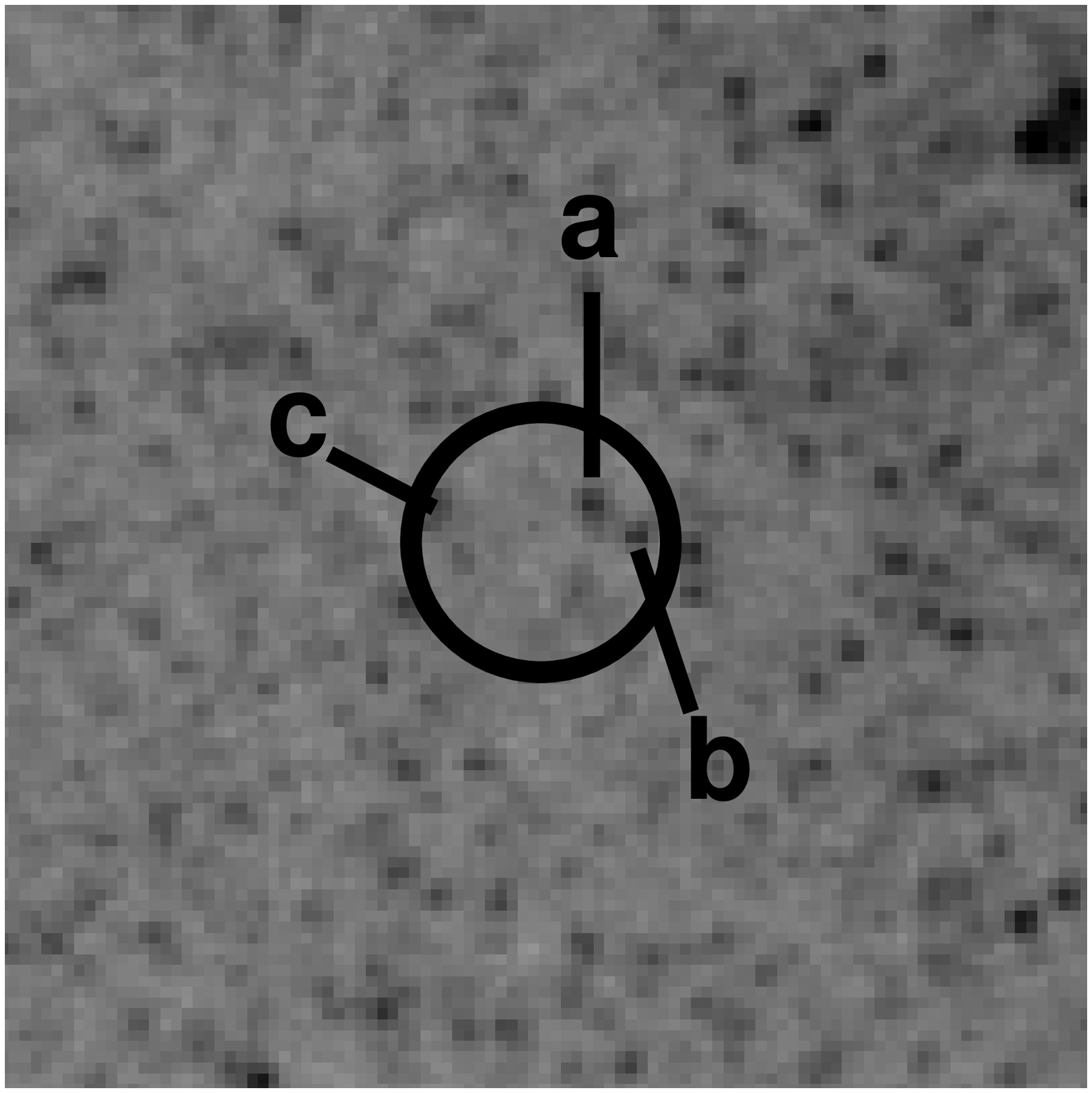} \\
T3 & T4 \\
\includegraphics[width=3.5cm,height=3.5cm,angle=0]{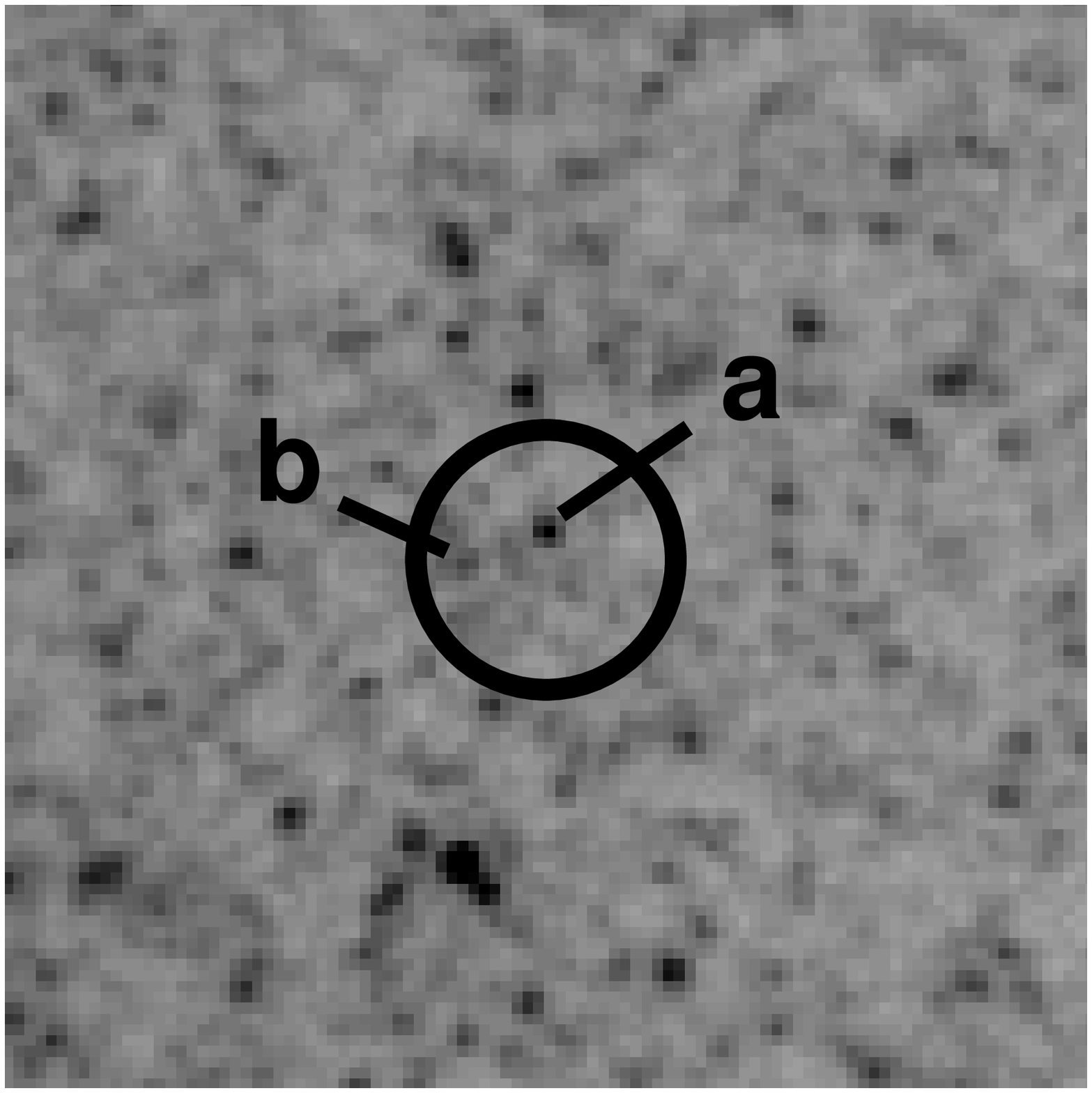} & 
\includegraphics[width=3.5cm,height=3.5cm,angle=0]{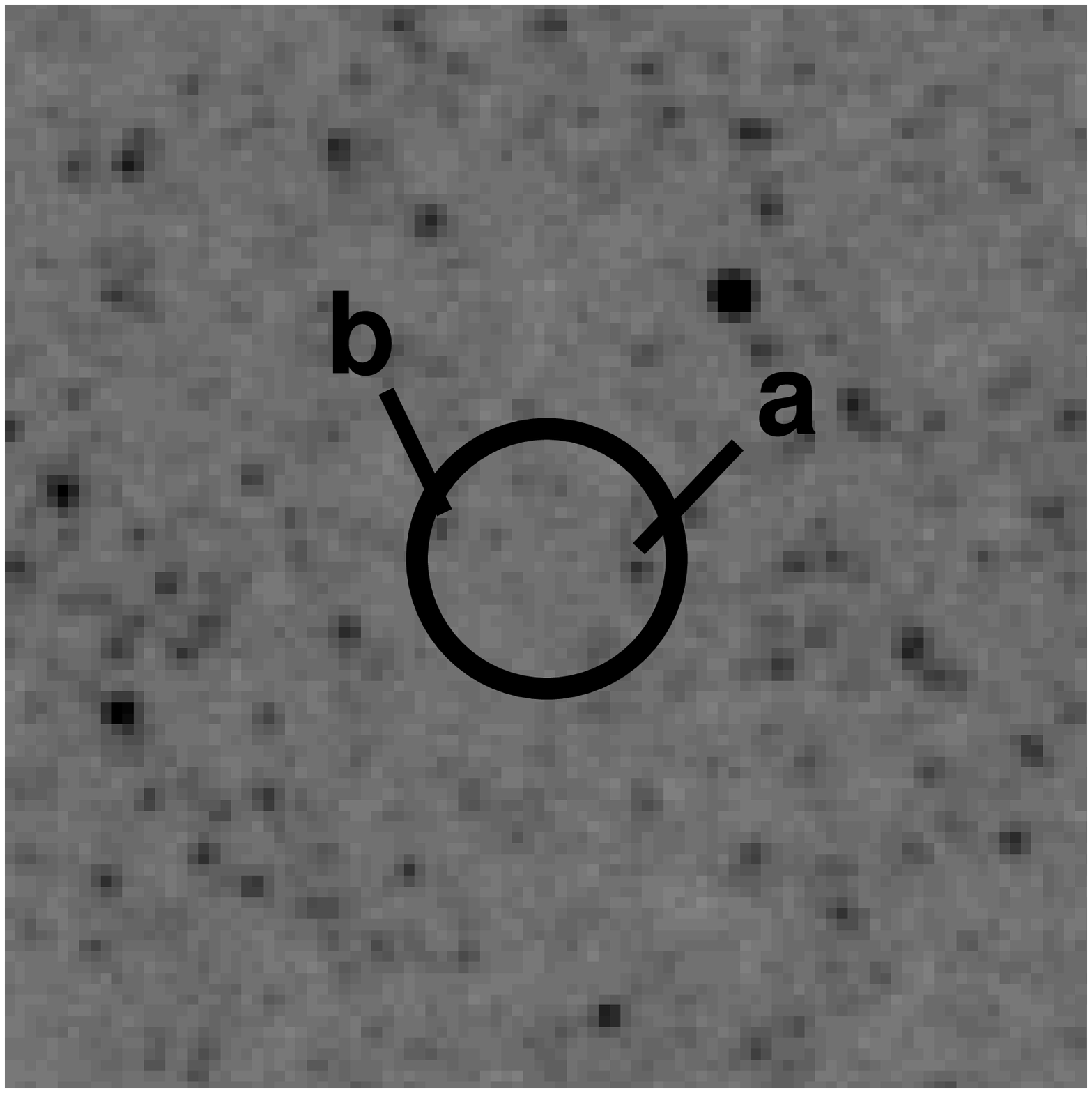} \\
T5 & T6 \\
\includegraphics[width=3.5cm,height=3.5cm,angle=0]{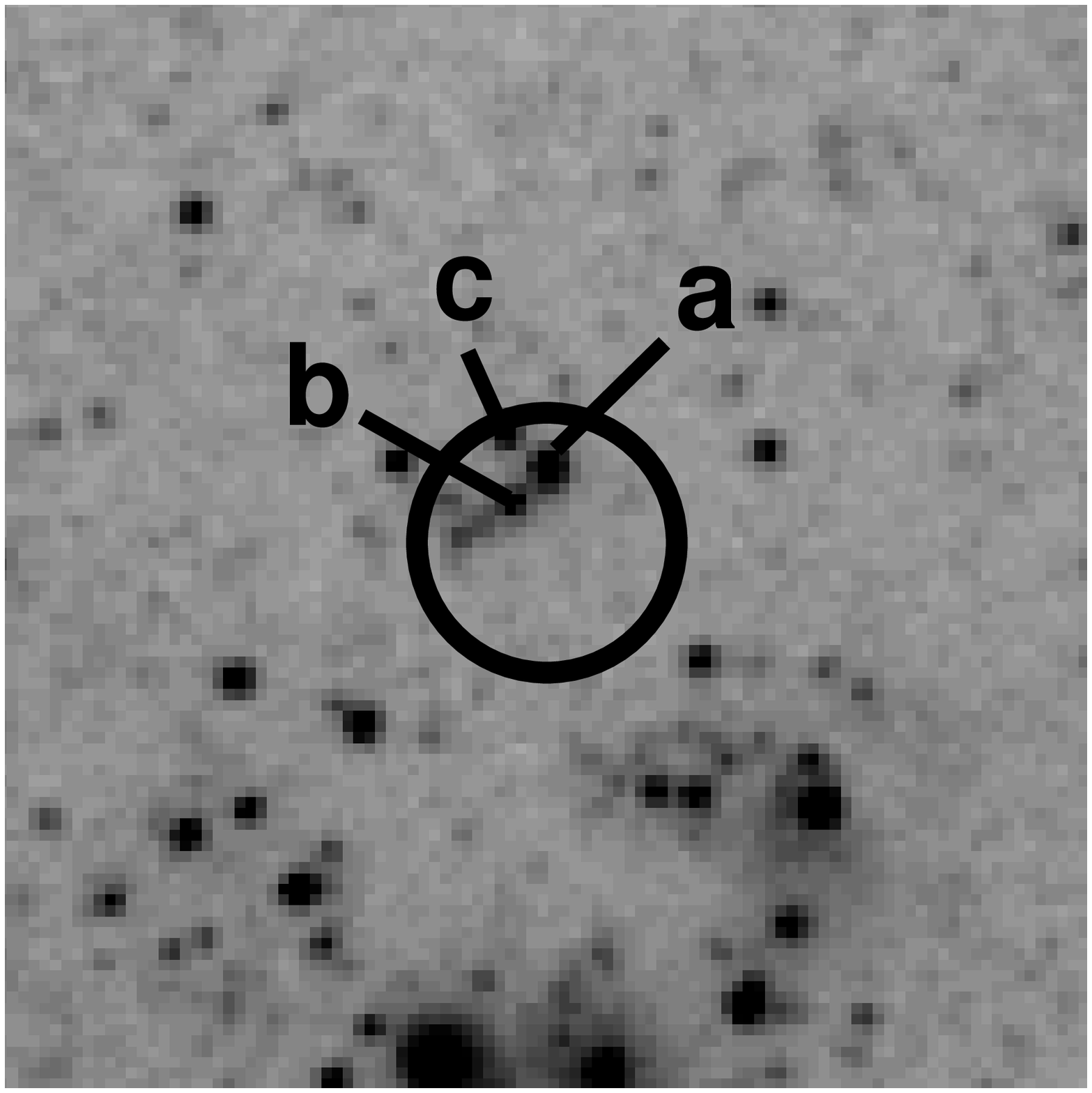} & 
\includegraphics[width=3.5cm,height=3.5cm,angle=0]{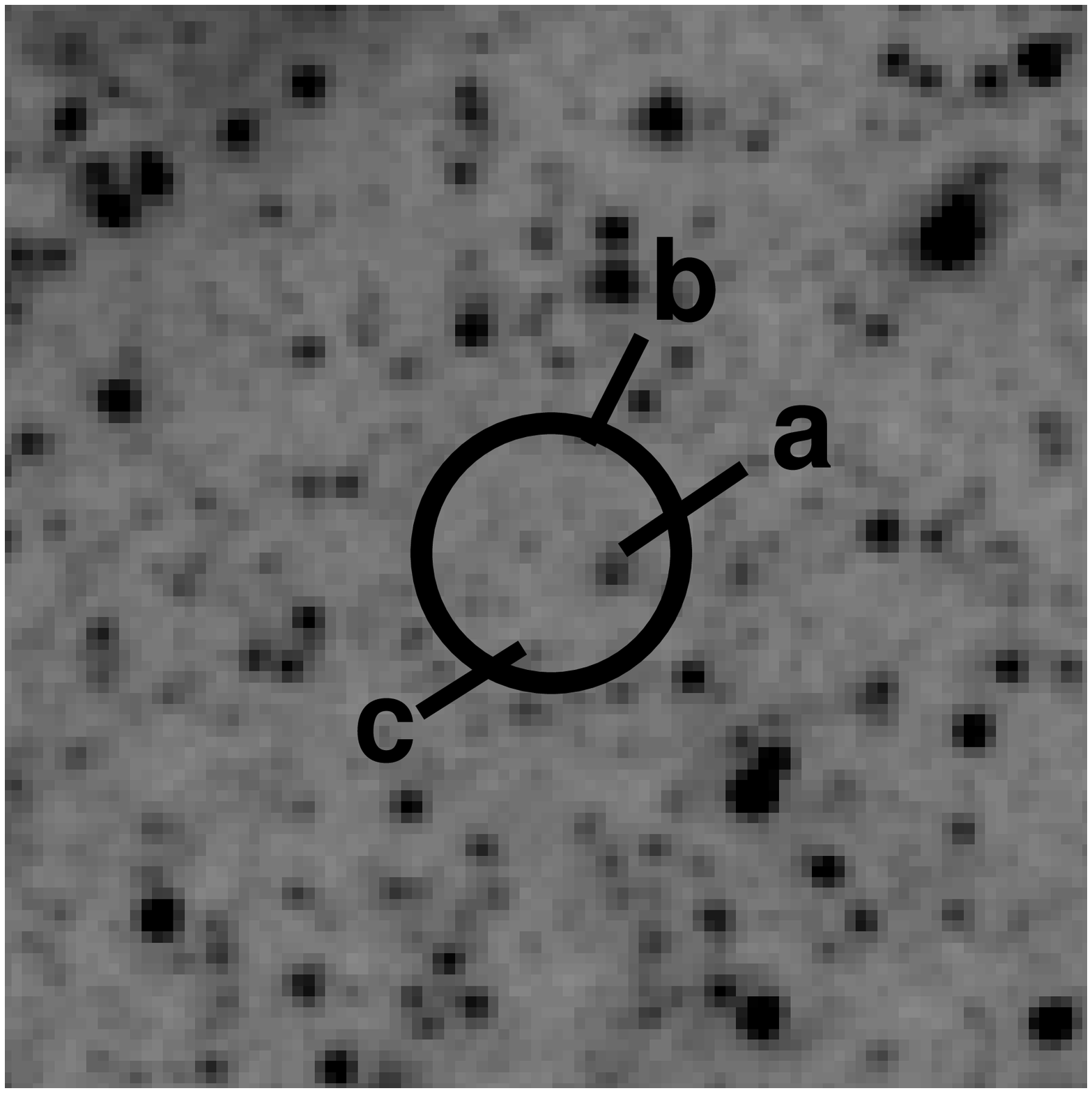}  \\
T7 &  \\
\includegraphics[width=3.5cm,height=3.5cm,angle=0]{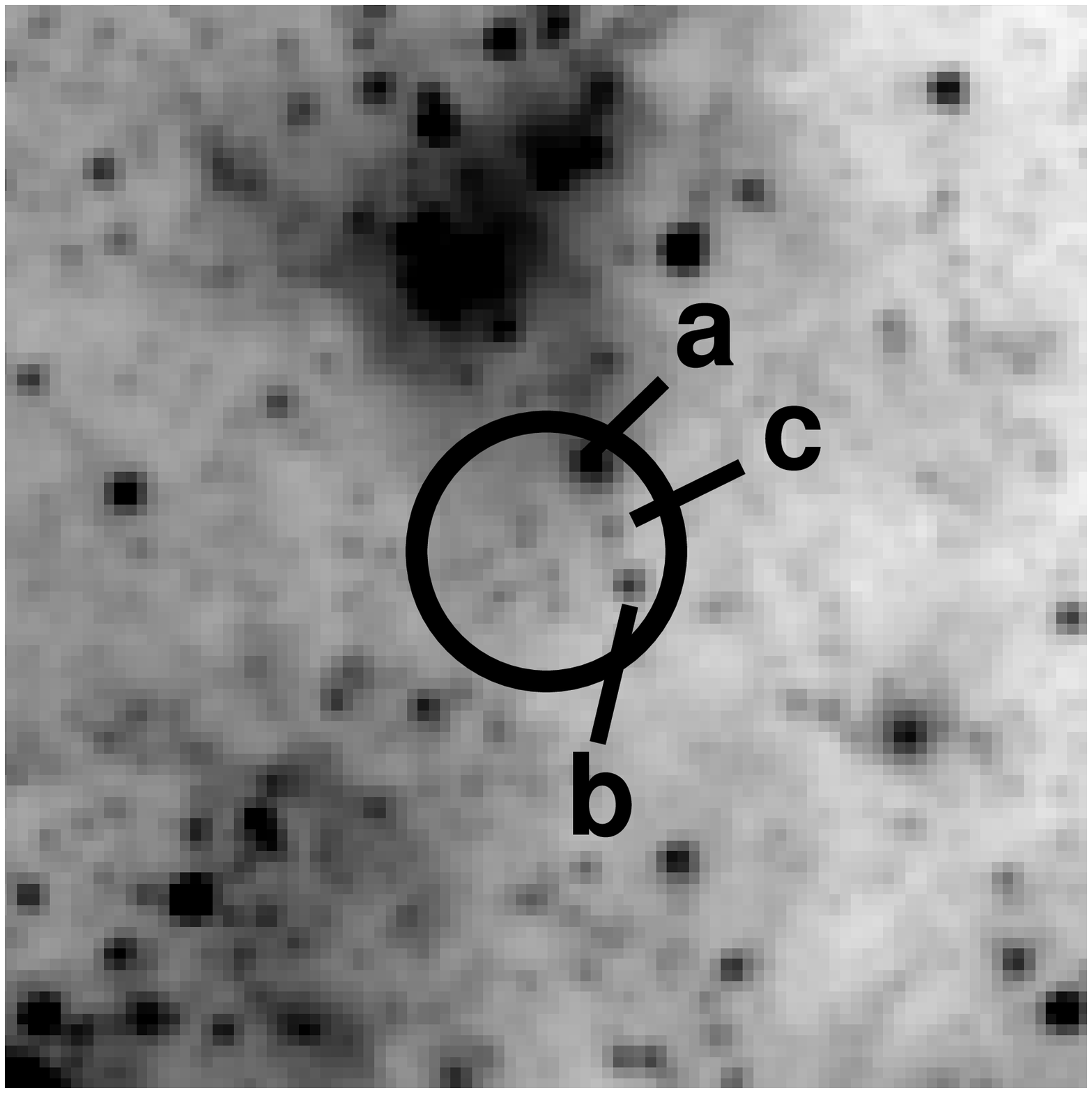} & \\

\end{tabular}
\caption{Optical \hst{} images for transient X-ray sources in NGC 4449. The box size is $5 \times 5$ arcsec. North is 
up, and east is to the left.The radius of the black circle is 0.6 arcsec. 
The three brightest objects inside 
the error circle were labeled as ``a'', ``b'' and ``c'' (or ``a" and ``b" when
there are only two sources) and 
their absolute magnitudes are given in Table \ref{optical}.}
\label{opt}
\end{figure}

The possible optical counterparts of the seven transient sources were 
searched in the $BVI$ band images obtained with \hst{} ACS. 
We performed the astrometric calibration of the X-ray and optical images 
using the stars from the U.S. Naval Observatory 
catalog \citep[USNO-B;][]{Mon03}. The plate solution was 
computed using the IRAF tool {\tt ccmap} and the root mean square (rms) 
residuals were typically less than a few tenths to hundredths of an arcsecond 
in both R.A. and decl. The total alignment errors were computed by 
summing X-ray and optical rms residuals in quadrature and applied 
to the coordinates of each X-ray source. The NGC 4449 POS-A and POS-B pointings 
covered six transients. While TSS1 is outside of the field of view of the both 
observations, the POS-C observation covered TSS1 in its field of view 
(close to the edge of the CCD). We identified the optical 
sources in each filter using IRAF {\sc daofind} task. Most of the transient 
source regions are 
crowded (see Figure \ref{opt}) and multiple optical sources were identified 
in the 0.6 arcsec radius error circle (90\% uncertainty of the 
\chandra{} X-ray absolute positions for 
ACIS-S\footnote{http://cxc.harvard.edu/cal/ASPECT/celmon/index.html}). 
The magnitudes of these optical sources were computed using 
the {\sc daophot} package in IRAF. We also calculated 
the logarithmic X-ray--to--optical flux ratio log$(f_{x}/f_{o})$, 
where $f_{x}$ and $f_{o}$ are the $0.3-8$ keV X-ray flux and 
the F555W flux respectively. For TSS1, we used the 
optical flux from the F814W band instead. The absolute magnitude, color, 
X-ray--to--optical flux ratio, and the likely class of each source are 
given in Table \ref{optical}.

\begin{table}
\tabletypesize{\small}
\tablecolumns{6}
\setlength{\tabcolsep}{0.2pt}
\tablewidth{320pt}
	\caption{Possible optical counterparts of transient X-ray sources in NGC 4449}
 	\begin{tabular}{@{}lccccr@{}}
	\hline
	\hline
\colhead{Src} & \colhead{Opt} & \colhead{$M$} & \colhead{Range of} & \colhead{Range of} & \colhead{log} \\
\colhead{No.} & \colhead{Src} &               & \colhead{$(B-V)$}  & \colhead{$(V-I)$}  & \colhead{$(f_{x}/f_{o})$} \\
\hline


		TSS1 & a & $-3.3 (I)$ & ...             & ...             & $[0.80, 1.43]$(H/A?) \\
		     & b & $-1.7 (I)$ &                 &                 &   \\
		TSS2 & a & $-1.5 (V)$ & $[-0.50, 0.86]$ & $[-0.29, 0.64]$ & $[0.31, 0.55]$(H) \\
		     & b & $-1.4 (V)$ &                 &                 &     \\ 
		     & c & $-0.9 (V)$ &                 &                 &     \\ 
		T3   & a & $-2.4 (V)$ & $[0.54, 1.51]$  & $[1.13, 1.21]$  & $[-0.27, -0.13]$(H) \\  
		     & b & $-2.1 (V)$ &                 &                 &     \\     
    		T4   & a & $-0.9 (V)$ &$[0.78, 1.54]$  & $[0.51, 0.98]$  & $[0.72, 0.86]$(H) \\
		     & b & $-0.5 (V)$ &                 &                 &     \\  				     
		T5   & a & $-3.3 (V)$ & $[-0.15, 0.01]$ & $[-0.50, -0.14]$ & $[-0.45, 0.01]$(H) \\
		     & b & $-2.6 (V)$ &                 &                 &     \\ 
		     & c & $-2.5 (V)$ &                 &                 &     \\
		     
		T6   & a & $-2.7 (V)$ & $[0.27, 1.33]$ & $[0.07, 1.78]$ & $[-0.23, 0.53]$(H) \\
		     & b & $-1.5 (V)$ &                 &                 &     \\ 
		     & c & $-0.8 (V)$ &                 &                 &     \\

		T7   & a & $-5.7 (V)$ & $[-0.50, 1.11]$ & $[-0.81, 1.88]$ & $[-1.65, -0.23]$(H) \\
		     & b & $-4.1 (V)$ &                 &                 &     \\ 
		     & c & $-3.6 (V)$ &                 &                 &     \\ 		
\hline
\end{tabular} 
\tablecomments {(1) Source number; (2) optical sources marked in Figure \ref{opt}; (3) the absolute magnitude of the marked optical sources and the corresponding filter is given in bracket; (4)-(5) range of $B-V$ and $V-I$ colors of the optical sources; (6) range of logarithmic X-ray--to--optical flux ratio and the likely class (H = HMXBs and A = AGN) is given in bracket.}
\label{optical}
\end{table}

X-ray--to--optical flux ratios can be used to identify the class of the X-ray 
sources. The flux ratios of stars have wide-range values 
of typically log$(f_{x}/f_{o}) < 0$, while those of 
the active galactic nuclei (AGNs) and BL Lac objects are
$>1$ \citep{Mac82, Sto91}. In SMC, the X-ray 
pulsars and high-mass X-ray binaries (HMXBs) have a flux ratio 
of $\lesssim 1$ with $B-V \lesssim 0$ \citep{Mcg08}. All the optical sources, 
except one source in the TSS1 error circle, have the flux 
ratio $< 1$. Thus, they are consistent with X-ray pulsars or HMXBs. In SMC, 
the Be-XRBs dominate the population of HMXBs with $M_{V} \sim -2$ 
to $-5$ \citep{Mcb08}. At the distance of NGC 4449, 
these absolute magnitudes correspond to $m_{V} \approx 25 - 22$, 
and the majority of the optical sources in the error circles are consistent 
with this range. Considering some of them as possible counterparts, they
would thus likely be Be-HMXBs. 
However, the colors of these optical sources are in the range of 
$B-V \sim [-0.5, 1.5]$ and $V-I \sim [-0.8, 1.9]$, 
which are broadly consistent with blue/red bright giants or supergiants. 
Among the transient sources, T7 \citep[X46 in][]{Ran11} was identified 
to be associated with a young star cluster of age $\sim 5$\,Myr. Our 
analysis also identified the star cluster 
in the T7 region (source ``a''), although there are other less bright 
sources in the region.

\section{Discussion}

Using archival \chandra{}, \xmm{}, and \swift{} observations,
we studied the transient X-ray sources in the Magellanic-type galaxy NGC 4449. 
We found seven transient sources, two of which are classified as 
SSSs because of their extreme soft X-ray colors, four are in the XRB class 
(T7 is considered as an XRB), and one source belongs to the absorbed class. 
These sources were also identified in the previous studies, 
but our analysis further classified them as transient X-ray 
sources in NGC 4449. 

\subsection{Transient SSSs}

The deep sensitivities of \chandra{} observations 
have indicated that the transient SSSs changed their fluxes 
by at least 1--2 orders of magnitude and become undetectable in
the later observations.
The spectra of the transient SSSs TSS1 and TSS2 were better described by 
an absorbed blackbody model, with the 
temperature ranging from $\sim 75$ to 105 eV. The X-ray luminosity 
in the $0.3-8$ keV band was 
$\sim 10^{38} {\rm~erg\ s}^{-1}$ for TSS1, while TSS2 was less luminous 
compared to TSS1, with $L_X\sim 10^{37} {\rm~erg\ s}^{-1}$. 
The effective blackbody temperatures and luminosities inferred
from the X-ray spectra are consistent with the characteristics of 
steady nuclear burning on $\sim 0.6-1.4~\Msun$ WDs \citep{Van92, Nom07}. 
Apart from this, the transient SSSs displayed 
short-term X-ray variability in the X-ray observations, which supports 
the binary nature of the systems. The supersoft emission at lower luminosities
can be from cooling isolated NS, Galactic stars, cataclysmic 
variables (CVs), and AGN \citep{Kah06b, Kah06, Kah08}. However, 
the transient behavior, binary nature, and high X-ray luminosity help 
to rule out the isolated stars, CVs, and AGN as the possible 
class for the transient SSSs. The X-ray luminosities inferred 
from \chandra{} and \xmm{} observations are also consistent with 
those of Be/NS X-ray binaries \citep{Rei11}. Be/NS systems 
dominantly emit hard X-ray radiation, but sometimes a soft excess is also 
expected along with hard X-ray emission \citep{Hic04}. For our cases,
only supersoft X-ray emission was seen, indicating that they unlikely are
Be/NS binaries. Thus, the transient SSSs may belong to the X-ray binary 
class (HMXB) with a WD as the compact object and supersoft X-ray emission 
is due to the stable nuclear burning 
on the WD surface. This possibility is supported by the results
from the optical analysis of the images of the source fields. 

In TSS1, we found possible periodic variations from one \chandra{} observation 
with a timescale of $\sim 2.3$\,h. Considering the timescale as the orbital 
period, it is actually consistent with those of the luminous SSSs in our 
Galaxy, SMC and LMC, derived from photometric variations \citep{Kah97}. 
Moreover, the post-nova phase of classical, recurrent or symbiotic 
nova is identified in several SSSs and their orbital periods are in 
the range of $\sim 85$ min to 5.3 h \citep{Kah06b}. However with 
the available data, we covered only a few cycles of the possible modulation,
and it is thus hard to clarify the periodic nature of the variability. In order 
to confirm it, future sufficiently long X-ray observations are required 
once it is back to be bright again.

TSS1 was detected with \einstein{}, \rosat{}, \chandra{}, and \xmm{} 
observations. Despite of the ill-sampled observations, the source was 
possibly active for $\sim 23$ years and the X-ray luminosity 
was nearly constant around $\sim 10^{38} {\rm~erg\ s}^{-1}$. 
Possibly after year 2002 or 2007 the source became undetectable, even in
deep \chandra{} observations.
These results suggest that TSS1 may belong to a sub-class of SSSs 
with a small orbital period ($\sim 1-5$ h), similar to classical novae, 
but with long SSS phase ($> 10$ years). 
Such characteristics favor a slightly low mass WD 
($\sim 0.6 - 0.8\,\Msun$) in the system \citep{Hac08, Kat10}. Moreover, 
TSS1 has seemingly shown a decreasing trend in temperature 
($\sim 105$ to 86 eV), which may provide the further support for a lower mass 
WD system \citep{Kah08}. TSS2 exhibited short supersoft X-ray phase (detected only in 2001 
observation), indicative of a massive WD system \citep{Kat10}.
 
The non-detection of X-ray emission from the transient SSSs can be due to the 
exhaustion of nuclear burning on the WD surface. Such scenario is possible 
for post-nova SSSs, if their accretion rate is too low to sustain the steady 
nuclear surface burning. A nova explosion (either classical or recurrent) 
in future from transient SSSs can re-ignite the nuclear burning, 
which is followed with another bright supersoft X-ray phase \citep{Stu12}. 
This feature gives importance to the future observations of the source field. 

The post-nova X-ray emission has been observed 
in many SSSs, for example the first discovered SSS Nova Muscae 1983 
(GQ Mus; \citealt{Oeg84}). GQ Mus has been observed to emit in soft X-ray for a 
long period ($\sim 10$ years; \citealt{Oeg93}) after outburst and later 
turned off due to the complete consumption of the residual 
material \citep{Sha95}. 
The nova explosion and SSSs connection has been well established by 
the frequent observations of the central region of 
M31 \citep{Hen10, Hen11, Hen14}. 
Using the dedicated observations in 2006 - 2012, 
a sample of 79 novae with SSS counterparts in M31 has been identified
and the duration of their SSS phase was determined.
Among the novae, M31N 1996-08b and M31N 1997-11a were active for more 
than a decade ($\sim 13.8$ and $\sim 11.5$ yr 
respectively) in X-rays after their optical discovery \citep{Hen11} and became 
undetectable in the later observations \citep{Hen14}. Since the outburst 
history is not known for TSS1 and TSS2, the turn-off time of these sources 
can be estimated as $> 25.9$ and $> 0.7$ yr respectively 
(by assuming the mid point between the \xmm{} and 
\chandra{} observations). TSS1 was detected with most of the X-ray 
observatories (see \S \ref{long_lc}), i.e., active for $\sim 23$ yr, 
thus one can argue that TSS1 has the longest SSS phase 
known so far followed by the Galactic nova V723 Cas \citep{Nes08, Sch11}, 
which has the SSS turn-off time $\sim$ 18--19 yr \citep{Nes15}. 

Some of the classical SSSs exhibited the X-ray off-state for a short timescale. 
For example, the prototypical SSS CAL83 underwent eight temporary X-ray 
off-states during 1996 -- 2008 \citep[see][and references therein]{Kah98, Gre02, Raj13}. All 
the X-ray off-states occurred during optical high states \citep{Raj13} and 
their duration is less than 120 d \citep{Kah98}. The most likely explanation 
for the observed X-ray and optical variability is the expansion/contraction of WD's 
photosphere \citep{Gre02}. In the case of TSS1 and TSS2, one can assume 
that the X-rays turned off after year 2002 and 2001 (the last detected X-ray observation) 
respectively, but they were not caught in its normal X-ray on-state in any of the 
later observations. The available observations could have missed 
the X-ray on-state and such variability can be confirmed by the continuous 
monitoring observations in X-ray and optical wavelength.

\subsection{NS and BH X-ray Transients}

Four less-luminous transient sources (T3, T4, T5 and T6) were fitted with 
an absorbed PL with $\Gamma \sim 0.4 - 2.4$ and their luminosities 
are $\leqslant 10\%$ Eddington luminosity ($L_{Edd}$) for NS and BH primaries. 
The NS or BH XRBs exhibit the hard-state \citep{Rem06} spectrum only at 
luminosities $\leqslant 10\% L_{Edd}$ \citep{Gla07}. 
Thus, the spectral properties indicate that these sources are possibly
NS or BH XRBs at the hard state in NGC 4449. Source T7 had a steep PL 
index from the spectral fit, which indicates its soft emission. 
The spectral modeling with blackbody model provided 
a temperature of $\sim 180$ eV, which is higher than that of the typical 
SSSs, but consistent with QSSs \citep{Dis04}. QSSs in external galaxies 
are associated with SNRs and NS or BH binaries \citep{Ori06, Dis10}. 
Given that the transient nature of T7 rules out the SNR classification,
a variable QSS, associated with a NS or BH X-ray binary, is a 
rather case for T7.

\subsection{Possible Optical Counterparts and Comparison With Magellanic-type Galaxies}

The possible optical counterparts of the transient sources were searched 
in \hst{} observations, and we found that the source regions contain 
multiple optical sources. Without information such as correlated 
variability, it is hard to identify whether these optical sources are 
the counterparts or not. Also we note that the counterpart of a post-nova 
system would be too dim to be detected with the \hst{} observations. In any case, 
the absolute magnitudes, optical colors and X-ray--to--optical flux ratios of 
the optical sources in the source regions are broadly consistent 
with those of HMXBs, if we assume one of the optical sources in each
error circle as the counterpart. In addition, the absolute magnitudes
of the majority of the optical sources identified in our analysis are 
consistent with the range of Be-HMXBs in SMC \citep{Mcb08}. 
In the case of SSSs, previous studies suggest that they may have high-mass 
companions depending on their location in the host 
galaxy \citep{Dis04, Kun05}, and in the nearby galaxies they are associated 
with early-type stellar systems \citep{Li12, Stu12}. Thus we can argue 
that TSS1 and TSS2 are likely WD X-ray binaries with early-type stellar 
companion, i.e., WD Be/X-ray binaries, in NGC 4449. However, such Be/WD X-ray 
binaries are rare and only a few systems \citep[XMMU J052016.0-692505 in LMC, 
XMMU J010147.5-715550 and MAXI J0158-744 in SMC;][]{Kah06a, Stu12, Li12} 
have been reported in the past. If confirmed, the two transient SSSs would
be the first Be/WD X-ray binary systems identified in NGC 4449.

Finally, the Magellanic-type galaxies like LMC, NGC 55 and NGC 4449 have 
remarkable similarities in their X-ray 
properties \citep[see Table 12 of][]{Sto06}. We further compared 
the properties of the transient X-ray sources in NGC 4449 with the 
known transient population of MCs and NGC 55 
\citep[see Table 7 of ][]{Jit16}. Although the number of the transients
in NGC 4449 is low, probably not a sufficient sample, the overall properties of 
the transient X-ray sources (fractions of HMXBs and SSSs, 
and their luminosities) are consistent with that of the other Magellanic-type 
galaxies. Our study has further confirmed that the transient X-ray sources in
NGC 4449 are also similar to those in MCs and NGC 55.

\acknowledgments

We thank the anonymous referee for the helpful comments that improved this manuscript. 
This research has made use of archival data of 
\chandra{}, \xmm{} and \swift{} observatories through the High Energy Astrophysics 
Science Archive Research Center (HEASARC) Online Service, provided by NASA Goddard 
Space Flight Center. VJ acknowledges the financial support from Chinese Academy of 
Sciences through President's International Fellowship Initiative (CAS PIFI, Grant 
No. 2015PM059). This research was supported by the Strategic Priority Research 
Program ``The Emergence of Cosmological Structures" of the Chinese Academy of 
Sciences (Grant No. XDB09000000) and the National Natural Science Foundation of 
China (11373055). Z.W. acknowledges the support by the CAS/SAFEA International 
Partnership Program for Creative Research Teams.


\end{document}